# Integrating Per-Stream Stat Tracking into Accel-Sim

Shichen (Justin) Qiao[*], Xin Su[*], Matthew D. Sinclair[+]

University of Wisconsin-Madison

* ECE Department + Computer Sciences Department

{sqiao6, xsu57}@wisc.edu ; sinclair@cs.wisc.edu

## 1. Summary

Accel-Sim is a widely used computer architecture simulator that models the behavior of modern NVIDIA GPUs in great detail [1]. However, although Accel-Sim and the underlying GPGPU-Sim model many of the features of real GPUs, thus far it has not been able to track statistics separately per stream. Instead, Accel-Sim combines statistics (e.g., cycles and cache hits/misses) across all simultaneously running streams. This can prevent users from properly identifying the behavior of specific kernels and streams and potentially lead to incorrect conclusions. Thus, in this work we extend Accel-Sim's and GPGPU-Sim's statistic tracking support to track per-stream statistics [2]. To validate this support, we designed a series of multi-stream microbenchmarks and checked their reported per-kernel, per-stream counts.

## 2. Background

Currently, Accel-Sim primarily utilizes `src/gpgpu-sim/gpu-cache.cc` to track cache statistics. The stats are stored in vector of vector of unsigned long long integers in `m_stats`, `m_stats_pw`, and `m_fail_stats`, where the outer vector represent access types and the inner vector represent access outcomes. For example, when the L1 or L2 cache is accessed, `inc_stats`, `inc_stats_pw`, or `inc_fail_stats` is called with an `access_type` and an `access_outcome`, and the corresponding vector entry is incremented by one.

This current approach potentially introduces inaccuracy to cache statistics when multiple streams are executed concurrently. For instance, when multiple streams are overlapped in Figure 1, if both streams access cache during the same cycle, the corresponding cache stats is only incremented once in total instead of once per stream. Moreover, it is also impossible to tell which kernel is updating a given statistics during the overlapping time from the current trace prints – making it hard to optimize the behavior of the underlying kernels. Although Accel-Sim allows users to serialize the streams such that they run in isolation (which enables them to obtain precise statistics for each kernel on each stream), since the streams are not running concurrently, the behavior would be different.

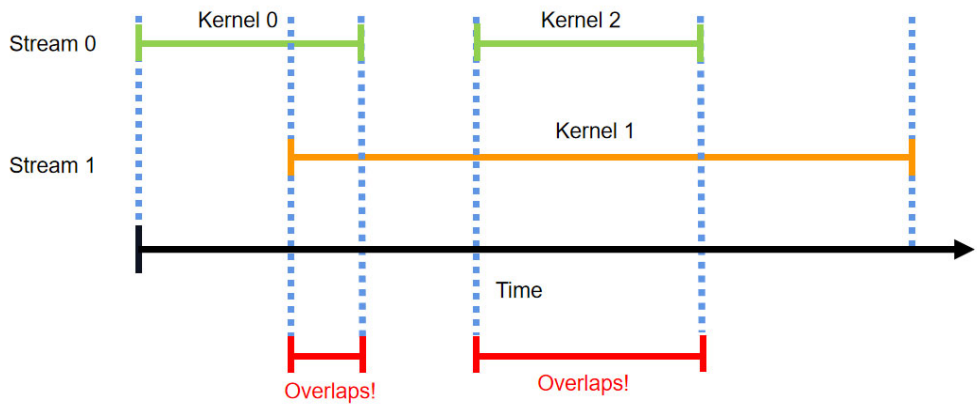

**Figure 1:** Example timing diagram with inaccuracy in cache stats caused by overlapping kernel executions.

# 3. Implementation

**Accel-Sim:**
Base Commit ID: 2260456ea5e6a1420f5734f145a4b7d8ab1d4737
Our Feature: https://github.com/accel-sim/accel-sim-framework/pull/166

**GPGPU-Sim:**
Base Commit ID: 13c67115070dc2f0876254a790d0238073ca364a
Our Feature: https://github.com/accel-sim/gpgpu-sim_distribution/pull/47

## 3.1 Cache Stats

To track statistics such as the cache stats in a per-stream fashion, we added an additional, outer dimension to the following data structures. For example, in `gpu-cache.h` we made the following change:

Before:
1. std::vector<std::vector<unsigned long long> > m_stats;
2. std::vector<std::vector<unsigned long long> > m_stats_pw;
3. std::vector<std::vector<unsigned long long> > m_fail_stats;

After:
1. std::map<unsigned long long, std::vector<std::vector<unsigned long long>>> m_stats;
2. std::map<unsigned long long, std::vector<std::vector<unsigned long long>>> m_stats_pw;
3. std::map<unsigned long long, std::vector<std::vector<unsigned long long>>> m_fail_stats;

Next we modified the constructor, mutators, operators, and accessors to reflect the above data structure change by adding a new, required parameter, `unsigned long long streamID`:

Before:
1. `void inc_stats(int access_type, int access_outcome);`
2. `void inc_stats_pw(int access_type, int access_outcome);`
3. `void inc_fail_stats(int access_type, int fail_outcome);`
4. `unsigned long long &operator()(int access_type, int access_outcome, bool fail_outcome);`
5. `unsigned long long operator()(int access_type, int access_outcome, bool fail_outcome) const;`

After:
1. `void inc_stats(int access_type, int access_outcome, unsigned long long streamID);`
2. `void inc_stats_pw(int access_type, int access_outcome, unsigned long long streamID);`
3. `void inc_fail_stats(int access_type, int fail_outcome, unsigned long long streamID);`
4. `unsigned long long &operator()(int access_type, int access_outcome, bool fail_outcome, unsigned long long streamID);`
5. `unsigned long long operator()(int access_type, int access_outcome, bool fail_outcome, unsigned long long streamID) const;`

`streamID` is already tracked in Accel-Sim in the `trace_kernel_info_t` class. Thus, it and can be accessed in Accel-Sim from an existing accessor function, `get_cuda_stream_id()`. However, to also access it in GPGPU-Sim, we modified the constructor of `trace_kernel_info_t` to also pass `cuda_stream_id` to `kernel_info_t` (which is accessible in GPGPU-Sim). In turn this allows us to determine which stream a kernel is running on at any places where a kernel object is used – which we used to propagated `streamID` to other data structures such as `mem_fetch` and `warp_inst_t` as well. Doing this allowed us to identify which stream a given statistic should be updating throughout GPGPU-Sim.

Moreover, the current Accel-Sim implementation would print all cache stats after any kernel exits. Thus, when multiple streams are executed concurrently and a kernel finishes, cache stats from other streams will also be redundantly printed. We modified this behavior to only print cache stats from the stream of the exiting kernel, by passing `cuda_stream_id` of that kernel to `print_stats` from the Accel-Sim side to GPGPU-Sim.

Before:
1. `void print_stats(FILE *fout, const char *cache_name = "Cache_stats") const;`
2. `void print_fail_stats(FILE *fout, const char *cache_name = "Cache_fail_stats") const;`

After:
1. `void print_stats(FILE *fout, unsigned long long streamID, const char *cache_name = "Cache_stats") const;`
2. `void print_fail_stats(FILE *fout, unsigned long long streamID, const char *cache_name = "Cache_fail_stats") const;`

### 3.2 Performance Stats

Next, to track the launch and exit times of each kernel on each stream we added the following data structures to `gpu-sim.h`:

1. `typedef struct {`

```
2.     unsigned long long start_cycle;
3.     unsigned long long end_cycle;
4.   } kernel_time_t;
5.   std::map<unsigned long long, std::map<unsigned, kernel_time_t>> gpu_kernel_time;
6.   unsigned long long last_streamID;
7.   unsigned long long last_uid;st;
```

These new fields are tracked and updated during `gpgpu_sim::launch` and `gpgpu_sim::set_kernel_done`, and they are printed out at the end of each kernel's statistics.

## 4. Usage

1. The "`-gpgpu_concurrent_kernel_sm`" flag should be set to 1 in the corresponding `gpgpusim.config` if per stream statistics are needed.
2. In `accel-sim-framework/`, set the environment by running:
```
source ./gpu-simulator/setup_environment.sh
```
3. In `accel-sim-framework/`, launch simulation by running (for example):
```
./gpu-simulator/bin/release/accel-sim.out -trace ./util/tracer_nvbit/traces/kernelslist.g -config ./gpu-simulator/gpgpu-sim/configs/tested-cfgs/SM7_TITANV/gpgpusim.config -config ./gpu-simulator/configs/tested-cfgs/SM7_TITANV/trace.config
```
4. In the output traces, locate "`Total_core_cache_stats_breakdown`" or "`L2_cache_stats_breakdown`" for detailed, per-stream cache statistics.

## 5. Validation

### 5.1 Verifying L2 Cache Stats

We first validated our design with a modified version of a simple benchmark with known, deterministic number of cache accesses, `l2_lat.cu` [1]:
https://github.com/accel-sim/gpu-app-collection/blob/release/src/cuda/GPU_Microbenchmark/l2_lat/l2_lat.cu

We chose `l2_lat.cu` because it has a deterministic number of L2 accesses, making it easy to verify. We modified it to use four streams of the same `l2_lat` workload in parallel:

```
1.   …
2.   #define THREADS_NUM 1    // one thread to initialize the pointer-chasing array
3.   #define WARP_SIZE 32
4.   #define ITERS 1           //iterate over the array ITERS times
5.   #define ARRAY_SIZE 1
6.
```

```
7.    __global__ void l2_lat(uint32_t *startClk, uint32_t *stopClk, uint64_t *posArray, uint64_t *dsink){
8.        // thread index
9.        uint32_t tid = threadIdx.x;
10.
11.       // initialize pointer-chasing array
12.       if (tid == 0){
13.           for (uint32_t i=0; i<(ARRAY_SIZE-1); i++)
14.   posArray[i] = (uint64_t)(posArray + i + 1);
15.           posArray[ARRAY_SIZE-1] = (uint64_t)posArray;
16.       }
17.
18.   if(tid < THREADS_NUM){
19.           uint64_t *ptr = posArray + tid;
20.           uint64_t ptr1, ptr0;
21.           …
22.           // pointer-chasing ITERS times
23.           // use cg modifier to cache the load in L2 and bypass L1
24.           for(uint32_t i=0; i<ITERS; ++i) {
25.               asm volatile ("{\t\n"
26.                 "ld.global.cg.u64 %0, [%1];\n\t"
27.                 "}" : "=l"(ptr0) : "l"((uint64_t*)ptr1) : "memory"
28.               );
29.               ptr1 = ptr0;   //swap the register for the next load
30.           }
31.           ...
32.       }
33.   }
34.
35.       int main(){
36.         ...
37.         cudaStream_t stream_1, stream_2, stream_3, stream_4;
38.         ...
39.         l2_lat<<<1,THREADS_NUM,0,stream_1>>>(startClk_g, stopClk_g, posArray_g, dsink_g);
40.         l2_lat<<<1,THREADS_NUM,0,stream_2>>>(startClk_g, stopClk_g, posArray_g, dsink_g);
41.         l2_lat<<<1,THREADS_NUM,0,stream_3>>>(startClk_g, stopClk_g, posArray_g, dsink_g);
42.         l2_lat<<<1,THREADS_NUM,0,stream_4>>>(startClk_g, stopClk_g, posArray_g, dsink_g);
43.         cudaStreamSynchronize(NULL);
44.         ...
45.       }
```

To validate our changes, we run three configurations: a) serialized kernel launches

(*tip_serialized*, which represents Accel-Sim without our changes and with streams serialized), b) without our changes (*clean*, which represents the baseline Accel-Sim without our changes), and c) with our changes under concurrent kernel launches (*tip*). Next we compared all cache statistics for these configurations.

Figure 2 shows the performance and cache results for these configurations. The timeline shows that we are now running the 4 kernels in concurrent fashion, and they take about the same amount of time because they are copies of the same kernel with the same access pattern. This is also reflected in the cache stats, which show that for the same access type and the same access results, the *clean* statistics always equals to the sum of our *tip* stats across all four streams. Note that when the kernels are serialized (blue bars), we may have more "HIT" counts compared to the concurrent *tip* counts (green bars). This is because the missing "HIT" counts under concurrent execution were counted as "MSHR_HIT" due to load dependencies among different streams. The total read and write access counts for each of the four streams are consistent and exactly met our expected counts.

**Figure 2:** l2_lat_4stream.cu validation results. Cache access counts with our new feature (green) exactly match their counterpart without our patches (orange).

Note that the serialized kernel executions mentioned above was achieved by applying this patch to Accel-Sim:

1. diff --git a/gpu-simulator/main.cc b/gpu-simulator/main.cc
2. index cbd42bd..fa6f1db 100644
3. --- a/gpu-simulator/main.cc
4. +++ b/gpu-simulator/main.cc
5. @@ -98,6 +98,9 @@ **int** main(**int** argc, **const char** **argv) {

```
  6.         std::cout << "Header info loaded for kernel command : " << commandlist[i].command_string << std::endl;
  7.         i++;
  8.       }
  9.  +   }
 10.  +   else if (kernels_info.empty())
 11.  +     assert(0 && "Undefined Command");
 12.
 13.       // Launch all kernels within window that are on a stream that isn't already running
 14.       for (auto k : kernels_info) {
 15. @@ -106,16 +109,13 @@ int main(int argc, const char **argv) {
 16.         if (s == k->get_cuda_stream_id())
 17.           stream_busy = true;
 18.       }
 19.  -    if (!stream_busy && m_gpgpu_sim->can_start_kernel() && !k->was_launched()) {
 20.  +    if (!stream_busy && m_gpgpu_sim->can_start_kernel() && !k->was_launched() && busy_streams.size() == 0) {
 21.         std::cout << "launching kernel name: " << k->get_name() << " uid: " << k->get_uid() << std::endl;
 22.         m_gpgpu_sim->launch(k);
 23.         k->set_launched();
 24.         busy_streams.push_back(k->get_cuda_stream_id());
 25.       }
 26.     }
 27.  - }
 28.  - else if (kernels_info.empty())
 29.  -   assert(0 && "Undefined Command");
 30.
 31.     bool active = false;
 32.     bool sim_cycles = false;
```

## 5.2 Verifying More Complex Microbenchmarks

After verifying the simple L2 microbenchmark in Section 5.1, we moved on to verifying more complicated microbenchmarks with concurrent streams. We chose these microbenchmarks because they are representative of applications which execute different kernels with different access patterns and behaviors on different streams. Thus, unlike Section 5.1, these microbenchmarks should exhibit more varied behavior across streams.

`benchmark_1_stream.cu` [5]:

```
  1.   __global__ void saxpy(int n, float a, float *x, float *y)
  2.   {
  3.     int i = blockIdx.x*blockDim.x + threadIdx.x;
```

```
4.      if (i < n) y[i] = a*x[i] + y[i];
5.    }
6.
7.    __global__ void scale(int n, float s, float *a)
8.    {
9.      int i = blockIdx.x*blockDim.x + threadIdx.x;
10.     if (i < n) a[i] = s*a[i];
11.   }
12.
13.   __global__ void add(int n, float *a, float *b)
14.   {
15.     int i = blockIdx.x*blockDim.x + threadIdx.x;
16.     b[i] = (i < n/2)? a[i] + b[i] : 2.0*b[i];
17.   }
18.
19.   int main(void)
20.   {
21.     ...
22.     cudaStream_t stream_1;
23.     cudaStreamCreate(&stream_1);
24.     …
25.     // Kernel 1
26.     saxpy<<<(N+255)/256, 256>>>(N, 2.0f, d_x, d_y);
27.     // Kernel 2 totally dependent on kernel 1
28.     scale<<<(N+255)/256, 256>>>(N, 2.0f, d_y);
29.     // Independent kernel 3
30.     saxpy<<<(N+255)/256, 256, 0, stream_1>>>(N, 3.0f, d_x, d_z);
31.     // Kernel 4: half its TBs are dependent on half of the threadblocks in kernel 2
32.     add<<<(N+255)/256, 256>>>(N, d_y, d_a);
33.
34.     cudaDeviceSynchronize();
35.     ...
36.   }
```

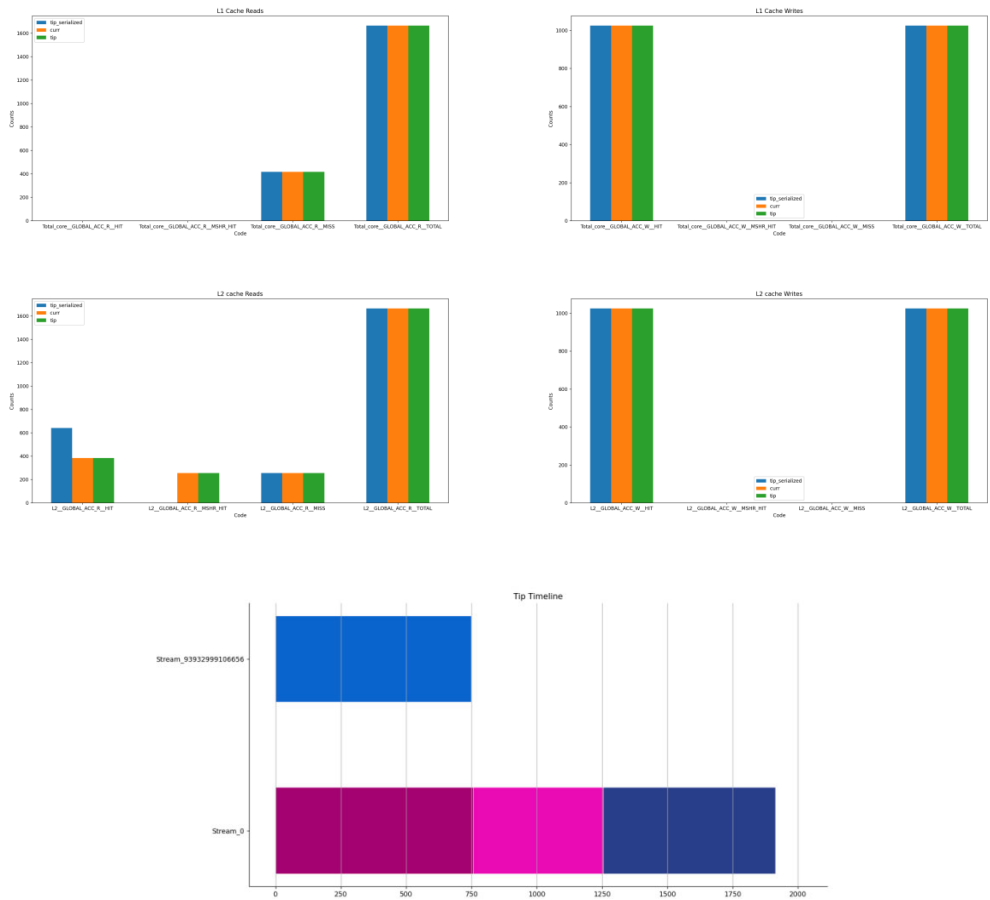

**Figure 3:** benchmark_1_stream.cu validation results. Cache access counts with our new feature (green) exactly match their counterpart without our patches (orange).

`benchmark_3_stream.cu` [5]:

```
1.  __global__ void saxpy(int n, float a, float *x, float *y)
2.  {
3.    int i = blockIdx.x*blockDim.x + threadIdx.x;
4.    if (i < n) y[i] = a*x[i] + y[i];
5.  }
6.
7.  __global__ void scale(int n, float s, float *a)
8.  {
9.    int i = blockIdx.x*blockDim.x + threadIdx.x;
10.   if (i < n) a[i] = s*a[i];
11. }
```

```
12.
13.   __global__ void add(int n, float *a, float *b)
14.   {
15.     int i = blockIdx.x*blockDim.x + threadIdx.x;
16.     b[i] = (i < n/2)? a[i] + b[i] : 2.0*b[i];
17.   }
18.
19.   int main(void)
20.   {
21.     int N = 1<<18;   //20
22.     ...
23.
24.     cudaStream_t stream_1;
25.     cudaStreamCreate(&stream_1);
26.      ...
27.
28.     // Kernel 1
29.     saxpy<<<(N+1023)/1024, 1024>>>(N, 2.0f, d_x, d_y);
30.     // Kernel 2 totally dependent on kernel 1
31.     scale<<<(N+1023)/1024, 1024>>>(N, 2.0f, d_y);
32.     // Independent kernel 3
33.     saxpy<<<(N+1023)/1024, 1024, 0, stream_1>>>(N, 3.0f, d_x, d_z);
34.     // Kernel 4: half its TBs are dependent on half of the threadblocks in kernel 2
35.     add<<<(N+1023)/1024, 1024>>>(N, d_y, d_a);
36.
37.     cudaDeviceSynchronize();
38.     ...
39.   }
```

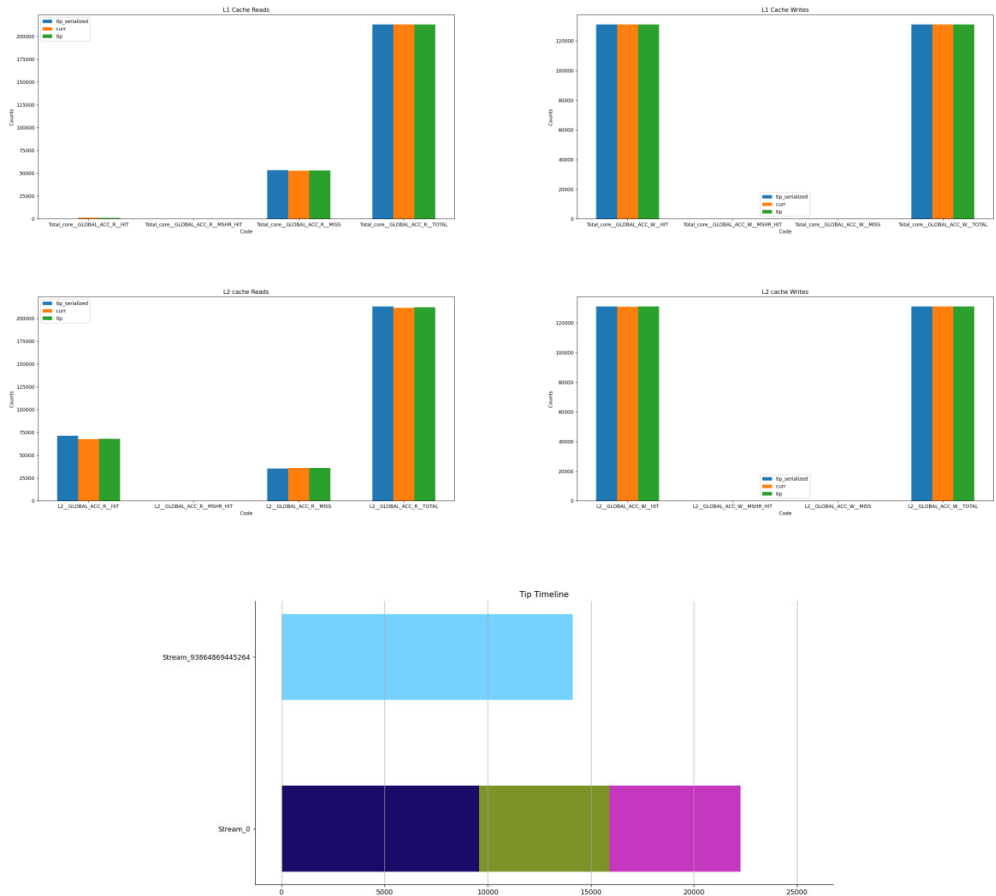

**Figure 4:** benchmark_3_stream.cu validation results. Cache access counts with our new feature (green) exactly match their counterpart without our patches (orange).

We did similar comparisons for `benchmark-1-stream.cu` and `benchmark-3-stream.cu`, and in Figures 3 and 4 we observe that, with our per-stream statistics tracking feature, the aggregated counts of a few access type/outcome combinations are increased: some green bars are slightly higher than their corresponding orange bars. These difference matches our expectation as described in Section 1 – sometimes when streams are running concurrently and stats are accessed in the same cycle from different streams, under-counting occurs. We also re-ran simulations of these two benchmarks multiple times and realized that the number slightly differ from run to run, due to determinism come from the benchmark source code. Still, in all scenarios, our counts are always greater than or equals to the stats came from the clean codebase, double confirming that, without our changes, Accel-Sim often under-count the cache stats.

## 5.3 Verifying DeepBench

Finally, we utilized a trace from DeepBench [3][4], inference_half_35_1500_2560_0_0, to further validate our implementations under more complex workloads with multiple streams (Figure 5). Unlike the previous microbenchmarks, DeepBench has many large kernels. Thus, it is difficult to validate precise counts for DeepBench. However, we nonetheless study it because it provides a sanity check that our changes do not significantly affect results in larger benchmarks.

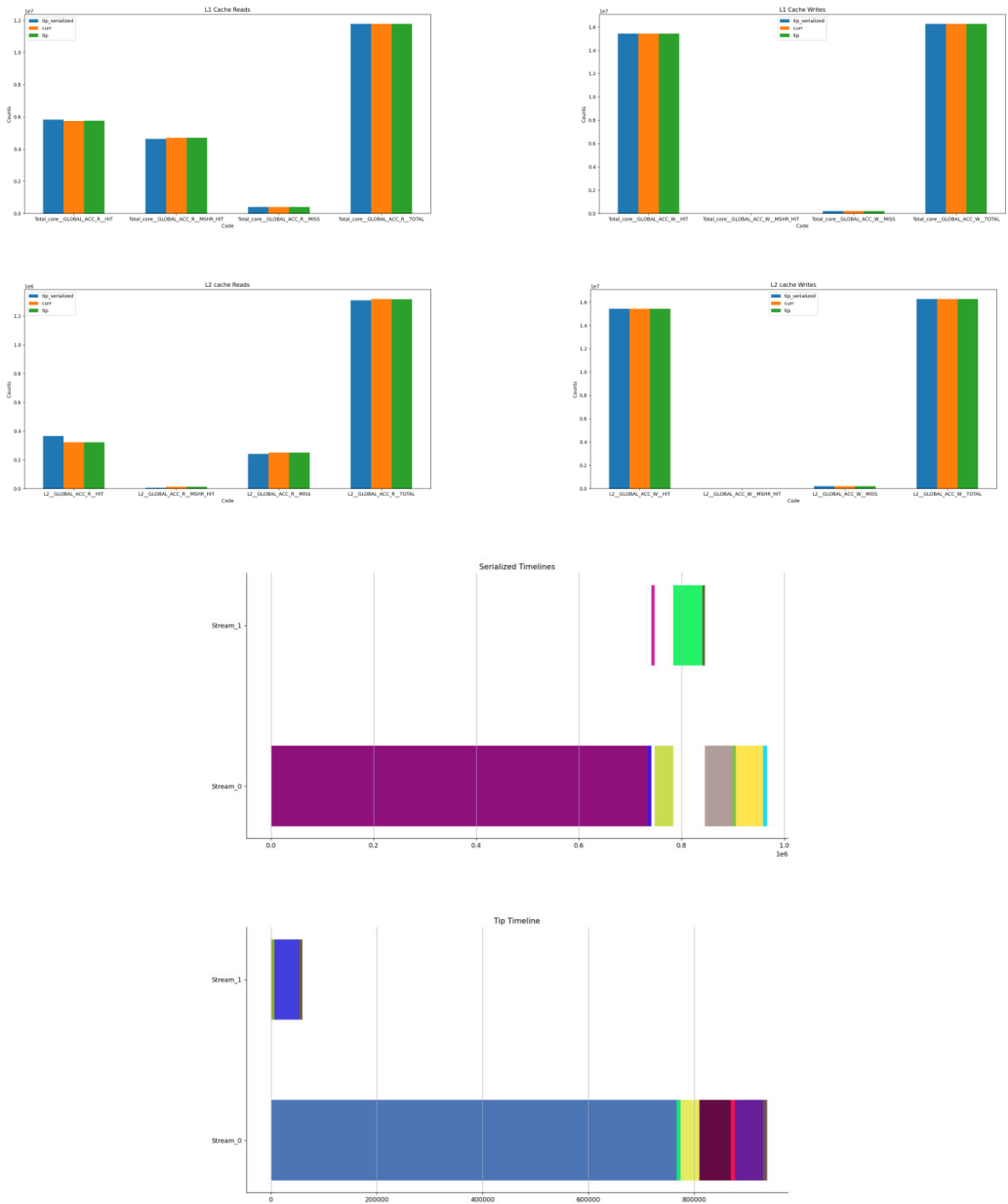

**Figure 5:** inference_half_35_1500_2560_0_0 validation results. Note that each kernel have different color in the timing diagrams.

Figure 5 shows the results for DeepBench's inference trace. Again, the behavior of our feature maintained in the same trends as they did during the simple benchmark validations documented above. Moreover, when looking at the performance timelines, we can see that our changes properly track the overlapping kernels – providing useful information that is not aggregated per cycle.

## 6. Limitations and Next Steps

Currently our support only tracks per-stream, per-kernel statistics for the caches (Section 3.1) and performance (Section 3.2). Nevertheless, since our aforementioned changes to pass `streamID` throughout GPGPU-Sim, similar feature expansions could also be developed for other components (e.g., interconnect, main memory).

In addition, since the `print_stats` function now require a `streamID` input argument, `power_stats.cc` and `gpgpusim_entrypoint.cc` could be affected. These modules are currently unaware of `streamID` and do not yet have access to the `get_streamID` functions we added. These imperfections should be revisited as needed once additional support is added.

## 7. Appendix: Graphing Tool

A graphing script was used to generate Figures 2 through 5. If more validation is needed, please select the desired stats to plot in the main function and execute the script with a command like this:

```
py .\graph.py .\l2_lat_4stream_tip_serialized.log .\l2_lat_4stream_clean.log .\l2_lat_4stream_tip.log
```

The graphing script and data used in this report is available here:
https://github.com/hal-uw/accel-sim-multiStream-stats

## 8. References

[1] M. Khairy, Z. Shen, T. M. Aamodt, and T. G. Rogers, "Accel-Sim: An Extensible Simulation Framework for Validated GPU Modeling," in 2020 ACM/IEEE 47th Annual International Symposium on Computer Architecture, ser. ISCA, 2020, pp. 473–486

[2] Jonathan Lew, Deval Shah, Suchita Pati, Shaylin Cattell, Mengchi Zhang, Amruth Sandhupatla, Christopher Ng, Negar Goli, Matthew D. Sinclair, Timothy G. Rogers, Tor M. Aamodt Analyzing Machine Learning Workloads Using a Detailed GPU Simulator, arXiv:1811.08933.

[3] Sharan Narang. DeepBench. https://github.com/baidu-research/DeepBench, 2016.

[4] Sharan Narang and Greg Diamos. An update to DeepBench with a focus on deep learning inference. https://svail.github.io/DeepBench-update/, 2017.

[5] Improving GPU Utilization in ML Workloads Through Finer-Grained Synchronization. Reese Kuper, Suchita Pati, and Matthew D. Sinclair. At 3rd Young Architects Workshop (YArch), April 2021.